\documentclass[a4paper,11pt]{article}
\usepackage[cp1251]{inputenc}
\usepackage[dvips]{graphicx}
\usepackage{epsfig}
\begin{document}
\Large



{\Large \textbf{
  \uppercase{Optical Multicolor $WBVR$--Observations of the X--Ray Star V1341~Cyg~$=$~Cyg~X--2 in
1986--1992}}\\

\bigskip\bigskip {Sternberg Astronomical Institute, Moscow State University,
Universitetskii pr. 13, Moscow, 119992 Russia}

\textbf{2010  A.N. Sazonov}

\vspace{12pt}

 \begin{abstract}
We present the results of $WBVR$ observations of the low--mass
X-ray binary V1341~$\textrm{Cyg} = \textrm{Cyg}$~X--2. Our
observations include a total of 2375 individual measurements in
four bands on 478~nights in 1986--1992. We tied the comparison and
check stars used for the binary to the $WBVR$~catalog using their
$JHK$~magnitudes. The uncertainty of this procedure was 3$\%$ in
the $B$ and $V$~bands and 8$\%$--10$\%$ for the $W$ and $R$~bands.
In quiescence, the amplitude of the periodic component in the
binary's $B$ brightness variations is within
$0.265^{m}{-}0.278^{m}$ ($0.290^{m}{-}0.320^{m}$ in $W$); this is
due to the ellipsoidal shape of the optical component, which is
distorted with gravitational forces from the X--ray component.
Some of the system's active states (long flares) may be due to
instabilities in the accretion disk, and possibly to instabilities
of gas flows and other accretion structures. The binary possesses
a low--luminosity accretion disk. The light curves reveal no
indications of an eclipse near the phases of the upper and lower
conjunctions in quiescence or in active states during the observed
intervals. We conclude that the optical star in the close binary
V1341~$\textrm{Cyg} = \textrm{Cyg}$~X--2 is a red giant rather
than a blue straggler. We studied the long-term variability of the
binary during the seven years covered by our observations. The
optical observations presented in this study are compared to
X--ray data from the Ginga observatory for the same time
intervals.
\end{abstract}


\newpage

\section{INTRODUCTION}

The X--ray source Cyg~X--2 was identified with a variable star,
later named V1341~Cyg~[1]. Cyg~X-2 is a low-mass, close binary
system with an accreting neutron star, and is one of the brightest
objects of this type in the X-ray.

According to its rapid X-ray variability and spectral behavior,
Cyg~X--2 belongs to the Z~class of X--ray sources~[2],
characterized by their Z-shaped tracks in the X-ray color--color
diagram. The Z-shaped track is traditionally subdivided into the
horizontal branch (HB), the normal branch (NB), and flaring branch
(FB), which correspond to the upper, middle, and lower parts of
the diagram, respectively.

It is generally thought that the position along the Z-shaped track
is related to the rate of matter accretion onto the neutron
star~[2--5], which is reflected in many optical photometric
features~[6--9].

The orbital period of V1341~Cyg was determined spectroscopically,
from the radial velocity of the X-ray binary's optical
component~[10, 11]. The orbital period is $P=9.8431^{\textrm{d}}$,
the epoch of periastron passage is $T_{0} = 2443161.7$, the
orbital eccentricity is $e=0.07$, the semi-amplitude of the radial
velocity curve is $R_v =87 \pm 3$~km/s, and the spectral type of
the optical component is F2III--IV.

Until recently, it was believed that the close binary
V1341~$\textrm{Cyg} = \textrm{Cyg}$~X-2 is characterized by two
physical states with considerably different optical light curves:
a quiescent and active state. The system displays two types of
activity.

1. Short-term flares, less than a day in duration, when the
combined brightness of the binary can increase by $0.85^m$ in the
$W$~band and by $0.75^m$ in the $B$~band. Brightness variations
within an observing night (about four to six hours of continuous
observations) are observed fairly frequently during periods of
activity of this type, followed by a period of smooth or rapid
fading (over 1.5--2~hours). The star is in quiescence on the
nights preceding and following the flare.

2. Long-term flares (some such active states of the binary are
probably due to instabilities in the accretion disk), five to ten
days in duration (or even longer), when the brightness increases
by $0.45^m{-}0.65^m$ in the $W$~band and by $0.35^m{-}0.45^m$ in
the $B$~band. In these cases, the level of the ellipsoidal light
curve quickly increases, stays at its maximum for three to four
days, and then smoothly decreases. This kind of activity sometimes
occurs cyclically~[12], but with cycles sometimes not being
observed for two to three months.

The X-ray observations of late 1980s and early 1990s (EXOSAT,
OSO-8, HEAO-1, Einstein, Ginga) considerably improved the existing
models of X-ray binary systems (see, for example,~[7, 13--16]).

The light curves we obtained for the system in 1986--1992 during
quiescent states qualitatively agree with the optical light curves
of other observers. There is a partial correlation with the X-ray
data for the same observing periods~[2].

The system contains a low-luminosity accretion disk, similar to
those of dwarf novae, and some of the system's active states
(long-term flares) may be due to instabilities of this disk.

It has been suggested that the evolutionary state of the close
binary V1341~$\textrm{Cyg} = \textrm{Cyg}$~X-2 makes it a blue
straggler~[12]. It was adopted in [12] that $E_{B{-}V}=0.22^{m}$,
and the UV excess was $B{-}V= 0.43^{m}$, $U{-}B={-}0.2^{m}$~[17];
Lyuty and Syunyaev~[17] also note rapid chaotic brightness
variations observed in the $B$~band in the range
$14.4^{m}{-}15.8^{m}$. Data on variations of the spectrum within
A5--F2 were presented in~[12]. These spectral variations occur
with the orbital phase, explained as a result of heating of the
optical star's surface by X-rays from its companion. The periodic
components in the light curve of the close binary could be due to
the ellipsoidal shape of the optical star (as for the close binary
V1357~$\textrm{Cyg} = \textrm{Cyg}$~X--1), the reflection effect,
or the heating of part of the optical star's surface by X--rays
from the companion, eclipses (as in the case of
V1343~$\textrm{Aql} = \textrm{SS}$~433), or a combination of
several of these effects (as for HZ~$\textrm{Her} =
\textrm{Her}$~X-1, where both the reflection effect and eclipses
are present). In this study, we have used refined values for the
mass of the neutron star and the radius of the Keplerian orbit, as
well as new estimates of the distance to Cyg~X-2 in the
relativistic precession model~[18], corresponding to the different
states of neutron stars according to~[19]. Taking into account
these serious corrections and the analysis of our many-year
photometry, we have good reasons to believe that Cyg~X--2 is not a
blue straggler.

We present here our long-term, multicolor optical observations of
the binary V1341~$\textrm{Cyg} = \textrm{Cyg}$~X--2.

\section{OBSERVING TECHNIQUES}

 Our photoelectric observations of V1341~$\textrm{Cyg} = \textrm{Cyg}$~X--2 were
performed in the $WBVR$ system. The observations comprise a total
of 2375~individual measurements on 478~nights during 1986--1992.

The ultraviolet $W$~filter ($\lambda _{\textrm{eff}} \approx
3500$~\AA, $\lambda _{1/2} \approx $ 520~\AA) is a revised version
of the standard $U$~filter [20].

Detecting periodicity in the variable's brightness variations
encounters certain difficulties because of the presence of a
large-amplitude irregular component in the brightness changes of
Cyg~X-2.

We monitored the object during a long time interval on each
observing night, obtaining observations from near the meridian to
an elevation corresponding to an air mass of $M(z)=1.5$.

The studies~[21, 22] were initially aimed at compiling all the
observations taking into account correction coefficients, the
response curves of the detecting equipment
(reflector~$+$~photometer), and systematic deviations between long
uniform series of observations for each season. The uncertainty in
the collected observations varies in the range
$0.04^{m}{-}0.06^{m}$ on different observing nights. There is
substantial scatter in the $W$ and $B$ bands. The scatter in the
$V$ and $R$~bands is somewhat smaller (the F star is quiescent).

Analysis of the light curves for the close binary
V~1341~$\textrm{Cyg} = \textrm{Cyg}$~X-2 requires long series of
high-quality uniform photoelectric observations in the
$W(U)BVRI$~bands~[10, 12, 17, 21], and their comparison to X--ray
observations from the same epochs~[23, 24].

\begin{table*}[t!]



\caption{Transformation coefficients}
\begin{tabular}{l|c|c|c|c|c}
\hline
\multicolumn{1}{c|}{Observing season} & $\xi_{V}$ & $\xi_{W{-}B}$ & $\xi_{B{-}V}$ & $\xi_{V{-}R}$ & $n$ \\
 \hline
July--September 1986, 1987, 1988 & $0.054~\pm$&0.997&0.929&1.068&27\\
(AZT-14 reflector (480 mm); Tien Shan &$\pm~0.002$&0.009&0.005&0.008&\\
High-Altitude Observatory, SAI)   &&&&&\\
\hline August--October 1986, 1987, 1988, &0.013&0.962&1.102&1.088&38 \\
1989, 1990, 1992 (Zeiss-600 reflector; &0.003&0.005&0.003&0.004&\\
Crimean Laboratory, SAI)&&&&&\\
\hline July--September 1986, 1987, 1988, 1990, 1992, & 0.012&0.958&0.937&1.065&41\\
 1994 (Zeiss-600 reflector; Mt. Maidanak  &0.003&0.004&0.007&0.007&\\
High-Altitude Observatory, SAI)&&&&&\\
\hline

\end{tabular}
\end{table*}

The shortest signal integration time during measurements of the
binary's brightness in the $W$, $B$, $V$, and $R$ bands was about
60--90~s (for studies of rapid variations). When measuring the
comparison stars, the check star, and the sky background, we
integrated the signal for about 60--120~s.

During each observing season, we also measured the transformation
coefficients between our instrumental photometric system and the
$WBVR$~system, as it was done in~[25]. We determine these using
the best photometric nights. We then averaged the
$\xi$~coefficients within one observing season and used the mean
$\xi$ values to calculate the zero points $\eta$ for each night of
the season. The resulting mean $\xi$~values and their
uncertainties are collected in Table~1, where $n$~denotes the
number of nights used to derive them. The same detecting equipment
was always used with the Zeiss--600 and AZT--14 reflectors.

To get a richer series of observations for the variable, we
observed the comparison star and the sky background each
30--40~minutes, interpolating to the time of observation of the
variable star. The particular instrumental system for the
observations presented here differs only insignificantly from the
standard $WBVR$ photometric system, so that the relations between
them can be represented with the linear equations




$$
\begin{array}{c}

V=v_0+\eta_v+\xi_v\cdot (B-V)\\
U-B= \eta_{(U-B)}+\xi_{(U-B)}\cdot(u-b)_0\\
B-V= \eta_{(B-V)}+\xi_{(B-V)}\cdot (b-v)_0\\
V-R=\eta_{(V-R)}+\xi_{(V-R)}\cdot (v-r)_0\\

\end{array}
$$

where the transformation coefficients $\eta$ and $\xi$ are
unknown.

The light detector was an FEU-79 photomultiplier (with an S-20
multi-alkaline photocathode). We calculated the $\eta$ and $\xi$
transformation coefficients anew each time the photomultiplier was
replaced.

It is difficult to specify a standard spectral energy distribution
for peculiar stars like V1341~Cyg = Cyg~$\textrm{X-2}$. In such
cases, the iterative technique for correcting for atmospheric
extinction in fundamental multicolor photometric observations
suggested in~[26] can be used to reduce broadband
$WBVR$~measurements to their values outside the atmosphere with
uncertainties not exceeding $0.005^m$ in the $W$~band or $0.003^m$
in the other ($B$, $V$, $R$) bands for telescopes with a 1000~mm
aperture and for air masses up to $M(z)=2$ in the case of
high-altitude observatories ($h \geq 3000$~m).

We found the reduction coefficients to the standard photometric
system from repeated measurements of standard stars in the SA 107,
108, 111--113 areas~[27]. We reduced the comparison and check
stars for our close binary to the $WBVR$~catalog based on their
$JHK$ magnitudes~[28] (Table~2). The uncertainties in this
reduction were within $3 \%$ for the $B$ and $V$~bands and $6\%$
for the $W$ and $R$~bands.

\section{METHOD USED TO REDUCE THE V1341~Cyg~$=$~Cyg~X-2 SYSTEM TO THE
ALMA-ATA $WBVR$ CATALOG OF BRIGHT NORTHERN STARS}

During all the observing seasons, we performed photoelectric
photometry of the binary V1341~Cyg~$=$~Cyg~X-2 in the $WBVR$
instrumental photometric system. We had access to the magnitudes
of the variable V1341~Cyg measured by other observers only in the
$UBV$ bands (for instance,~[10, 17]). We encountered certain
difficulties during our data reductions related to the
transformation of the instrumental $WBVR$ system to the Johnson
photometric system.

 We attempted to reduce our variable close binary to the uniform, high-accuracy
$WBVR$ photometric catalog using $JHK$ magnitudes from the 2MASS
catalog. This requires uniform, high-accuracy local photometric
standards from the ``$WBVR$ Catalog of Bright Stars of the
Northern Sky''{} (further called the $WBVR$ catalog)~[29].

We have sufficiently accurate $B$ and $V$~magnitudes for the
variable star, check star, and comparison star from published
data. The $JHK$~magnitudes for these stars were taken from the
2MASS catalog.

Let us apply an empirical formula relating the magnitudes in one
photometric system to the magnitudes and color indices in the
other. Obviously, we need at least one color index for this
purpose~[30].

We will use several color indices~[31, 32], enabling our
transformation formulas for the magnitudes to be more precise. In
the first stage, we adopt transformation equations from the $W$,
$B$, $V$, $R$ magnitudes to the magnitudes in other bands in the
form of a second-order polynomial in $V$ plus a complete cubic
polynomial in the three color indices:




m$_{B}$-m$_{A}$=a$_{0}$$^{.}$V + ${f^3}$ $^{.}$(W-B,B-V,V-R),

where A and~B are the two catalogs used to select the stars in
common for the comparison. We derived this third--order polynomial
via a least-square fit.

We obtained the following quantities:


m$^{calc}_{B}$= m$_{A}$+f$_1$
  $^{.}$(CI$^{A}_{1}$,CI$^{A}_{2}$,CI$^{A}_{3}$,...),

(where CI$^{\textrm{A}} _{i}$ are various color indices in the
A~catalog and $f_1$ is a polynomial of various powers of the color
indices, with cross-terms taken into account),


$\Delta$ m$_{B}$= m$_{B}$-m$^{calc}_{B}$;

$\sigma$$_{m}$=$\frac1n$ $\sum$ $\Delta$ m$^{2}_{B}$

If

$\Delta$ m$_{B}$>4 $^{.}$ $\sigma _{m}$

for some stars, we excluded these stars from the comparison and
again derived the above transformation relation.

The next stage of our calculations is to find the non-linearity
equation, which we solve via a least--squares fit:


m$_{B}$-m$_{A}$-f$_1$

 $^{.}$(CI$^{A}_{1}$,CI$^{A}_{2}$,CI$^{A}_{3}$,...)=f$_2$

 $^{.}$$(m_{A}$,m$^{2}_{A}$,...).

Finally,


m$^{calc}_{B}$= m$_{A}$+f$_1$
$^{.}$(CI$^{A}_{1}$,CI$^{A}_{2}$,CI$^{A}_{3}$,...)+f$_2$
$^{.}$$(m_{A}$,m$^{2}_{A}$,...)+f$_3$ $^{.}$($\alpha$,$\delta$).

where $\alpha$, $\delta$ are equatorial coordinates.

The $WBVR$ magnitudes for the comparison and check stars, which
have similar spectral types, were taken from the Alma--Ata $WBVR$
catalog. The observed $\Delta W$, $\Delta B$, $\Delta V$, and
$\Delta R$ values for V1341~Cyg are the differences between the
magnitudes of the comparison star and the variable (differential
photometry). The uncertainties in the calculated magnitudes of the
variable star for the $W$ and $R$ bands were $8\%$ and $6\%$,
respectively. In these calculations, it is also necessary to take
into account systematic errors of the catalogs, which depend,
among other factors, on the celestial coordinates of the program
stars.

\section{OBSERVATIONS}

In the results of the observations presented here, each data point
on the light curve represents an average of 2--16 individual
measurements. For our observations with the Zeiss-600 telescope,
the rms errors of a single measurement during times of maximum
brightness estimated from the pulse statistics correcting for the
background were $0.050^{m}$ for $W$, $0.025^{m}$ for~$B$,
$0.020^{m}$ for~$V$, and $0.015^{m}$  for $R$.

 We reduced the observations according to a differential scheme.
The photometric elements were taken from~[10, 33]:


T$_{o}$ = 2443161.7 + 9$^{d}$.8431$^{.}$E

where $T_{0}$ is the time of the periastron passage.




\begin{tabular} {|p{20pt}|p{80pt}|p{40pt}|p{35pt}|p{35pt}|p{35pt}|p{35pt}|p{20pt}|}

\hline   & JD2400000+.& $\varphi$&
$W$& $B$ & $V$ & $R$ & n \\
\hline 1& 46615.3830 & 0.873 &  15.075 &  15.125
& 14.667 & 14.328 & 06 \\
\hline 2& 46616.4574 & 0.983 &  15.144 &  15.314
& 14.946 & 14.660 & 08 \\
\hline 3& 46618.4099 & 0.181 &  14.855 &  15.306
& 14.779 & 14.585 & 06 \\
\hline

\end{tabular}


\section{REFINED ORBITAL PERIOD}

An analysis of the (O--C) data for the seven-year time interval
covered by our observations (plus the earlier published data for
the past ten years) demonstrates that the primary minimum occurs
at phase 0.0 and that the refined orbital period coincides with
the earlier value, $9.8431^{\textrm{d}}$, within the (${\pm}
0.0001^{\textrm{d}}$), indicating, in turn, that the
brightness-variation elements are in agreement with our
observations and need no correction. These calculations were done
using the code developed by V.P.~Goranskii (the WINEF1 software
package) and kindly made available to the author for the
mathematical reduction of these observations.

\section{OBSERVATIONAL DATABASE}

The results of our optical observations of the close binary
V1341~$\textrm{Cyg} = \textrm{Cyg}$~X-2 are collected in an
electronic database available at the address
http://lnfm1.sai.msu.ru/$\sim$sazonov/~Cyg~X-2. Table~3 presents a
small part of this database to illustrate its structure.
 The columns of Table~3 contain the (1) heliocentric Julian dates
of the observations, (2) orbital phases, $\varphi$, (3)--(6) $W$,
$B$, $V$, and $R$ magnitudes, and (7) number of individual
measurements~$n$.

\section{OBSERVATION RESULTS \protect\\ AND THEIR INTERPRETATION}

The combined results of our studies of the object in 1986--1992 in
the $WBVR$ bands are shown in Fig.~1; these consist of 2375
individual measurements on 478~nights.

The optical light curves of the binary in its active and quiescent
states differ considerably "--- among other features, in an
additional contribution to the brightness from the accretion disk
during active or intermediate states of the system. This can be
noted from the maximum brightness amplitude near Min~II (Fig.~1).
The regular, cyclic, small-amplitude flares lasting several days
are probably due to interactions of gas flows in the binary that
originate from the companion with the outer parts of the accretion
disk around the compact relativistic object, as well as to
accretion instabilities.

Figure~2 shows the photoelectric $WBVR$ light curves of V1341~Cyg
averaged in orbital-phase bins with widths of 0.025.

The second type of activity in the system is rapid flares with
durations of less than one day; it is natural to relate these to
processes in the X-ray source (we observe oscillations rather that
the flickering that is characteristic of dwarf novae).

Figure~3 displays a typical example of a rapid flare of the second
type, observed for 60~minutes on JD~2446735, when the brightness
increased by $0.04^{m}$ in $W$, $0.03^{m}$ in $B$ and $V$, and
$0.02^{m}$ in $R$. The brightness deviations from the mean are
within $3\sigma$.

\section{REGULAR, ORBITAL BRIGHTNESS VARIATIONS}

On average, the amplitude of the regular, orbital brightness
variations of the close binary is within $0.90^{m}$ in the
$W$~band, $0.95^{m}$ in~$B$, $0.60^{m}$ in~$V$, and $0.60^{m}$ in
$R$.

It follows from our observations that the variable V1341~Cyg
continuously exhibits chaotic variations of its brightness in the
ranges (Fig.~1):

$14.30^{m}{-}15.45^{m}$ in $W$;

$14.80^{m}{-}15.35^{m}$ in $B$;

$14.20^{m}{-}15.10^{m}$ in $V$;

$14.10^{m}{-}14.75^{m}$ in $R$.

\noindent The close binary also exhibits optical flares observed
over $\Delta t=3{-}4$~hours with amplitudes in the ranges:

$0.50^{m}{-}0.60^{m}$ in $W$;

$0.55^{m}{-}0.65^{m}$ in $B$;

$0.35^{m}{-}0.55^{m}$ in $V$;

$0.35^{m}{-}0.50^{m}$ in $R$.

\noindent Such flares of the second type were recorded, for
instance, on JD~2446619 and JD~2447370.

 The scatter of our individual data points around the mean light
curve has amplitudes of $0.30^{m}{-}0.35^{m}$ in $W$, $0.25^{m}
{-} 0.30^{m}$ in~$B$, $0.20^{m} {-} 0.22^{m}$ in~$V$, and
$0.15^{m} {-} 0.20^{m}$ in $R$. The amount of scatter depends on
the orbital phase. This is especially evident for the $W$-band
observation, indicating that ellipsoidal variations are most
pronounced in the low, quiescent state of the close binary.

 We can see from the observed light curves that the width of the
primary minimum (orbital phase~0.5) in the $W$, $B$, $V$, and $R$
bands varies approximately between 0.05 and 0.06.

\section{ORBITAL VARIATIONS OF THE COLOR INDICES}

 We find in our analysis of the ellipsoidal light curves of
V1341~Cyg in quiescence in all four bands ($WBVR$) that broader,
less sharp minima are characteristic of orbital phases near
$\varphi= 0.00$ (the upper conjunction of the F~star, with the
binary's relativistic component in front of it), as compared to
phases near $\varphi= 0.50$ (the lower conjunction of the binary,
with the normal component in front, characterized by narrower and
sharper minima). This pattern was observed in 1986, 1988, and
1991. It is probably due to the considerable eccentricity of the
close binary's orbit.

 Our observations confirm the earlier findings of other studies of
a clear functional dependence of $U{-}B$ (in our case, $W{-}B$) on
the deviation $\Delta B$ from the quiescent ellipsoidal curve,
with only a weak dependence observed for $B{-}V$: both color
indices decrease with increasing $\Delta B$. A similar relation is
observed for the color indices and brightnesses plotted as
functions of the orbital phase (Fig.~4).

 The object becomes redder when it fades in the $V$ band, as is
visible in the $ (V{-}R){-}V$ diagram in Fig.~5.

 No significant variations of the object's position were detected
in the other color--magnitude diagrams, namely $(W{-}B){-}B$,
$(W{-}B){-}V$, $(W{-}B){-}R$, and $(V{-}R){-}R$.

 Considerable variations of the $W$, $B$, $V$, and $R$ magnitudes
with $\varphi$ (Fig.~1) and of the $W{-}B$ and $V{-}R$ color
indices with $B{-}V$ (Fig.~6) were detected for each of the
observing seasons; these are probably due to heating of the
F-star's surface facing the X-ray source (the reflection effect).
This effect is manifest as a wave in the color-index diagrams near
the primary minimum (the zero orbital phase of the binary).
Exceptions occur for the $W{-}B$ color index in very low states of
the object, when the mean brightness of the star becomes about
$0.2^m$ below the mean level of the quiescent light curve for the
system.

\section{MEAN LIGHT CURVES
 \protect\\
FOR THE SYSTEM IN QUIESCENCE}

 We used the photographic observations with an orbital-phase
increment of 0.025 from Table~2 of~[12] as a basis for
constructing the mean light curve of V1341~Cyg in quiescence. The
mean uncertainty of the magnitudes in these orbital-phase bins is
$0.025^m {-} 0.035^m$. We then selected $B$-band photoelectric
observations from among our data that deviated from the mean
photographic curve (first translated to an intensity scale, as was
done in~[12]) by no more than $0.10^m$.

 The observations in the $W$, $V$, and $R$ bands corresponding to
the times of the $B$-band observations were properly synchronized.
As a result of this averaging, selection of photoelectric data
points, and synchronization of the points in the other bands, we
obtained the mean light curves for the binary displayed in Fig.~2.

 Note that the average dependences of $W{-}B$, $B{-}V$, and $V{-}R$
on $\varphi$ (Fig.~4) do not show any significant variations due
to heating of the side of the binary's optical component facing
the X-ray source (the reflection effect), which could lead to the
presence of a wave in the color indices with a minimum near zero
phase. The $B{-}V$ color index does not vary within the errors
(Fig.~4).

 The dependence of $W{-}B$ on $\varphi$ shows a double wave at the
orbital period and an amplitude of $0.4^m$. The lowest values of
this color index are observed at conjunctions. The same pattern is
detected for the original individual observational data points.

 Variations of the lower level (the lower envelope of the light
curve) in the scatter of the individual points in the $W$ light
curve are clearly visible. The upper envelope of the light curve
for the individual data points depends on the orbital phase. At
the upper level of the scatter of the individual points in the
$W$~band (and, partially, in the $B$~band), we can see a clear
wave with the orbital period of the binary. The maximum of the
wave is at orbital phase 0.85 (Fig.~1). Note that the wave with
the orbital period during the system's active state has no
functional relationship to the reflection effect or the appearance
of a hot spot on the accretion disk around the relativistic
object.

 Figure~7 displays the behavior of V1341~Cyg in its quiet and
active phases. We also performed similar analyses of the system's
behavior using the technique of Goransky and Lyuty~[12], who also
calculated the deviations $\Delta B$ from the ellipsoidal curve,
with the subsequent subtraction of the light curve drawn through
the distribution maximum of the individual data points from the
entire set of observations. We obtained approximately the same
result.

\section{RAPID VARIATIONS}

 We also studied rapid variations of the star on time scales
${\sim}60{-}90$~s in the $WBVR$ bands during its brightness maxima
and minima, during two hours of observations (Fig.~3).

 The flickering exhibited by the system has a very low amplitude,
and is suppressed by the presence of the bright, F-giant secondary
component.

 Our analysis of rapid variations indicated the presence of
brightness variations from the mean within $5\sigma$ on time
scales of 30~s. At those times, the optical component was at its
brightness maximum, namely:

1. JD 2447677; $W=14.390^{m}$, $B=14.833^{m}$, $V=14.338^{m}$,
$R=14.111^{m}$ at $\varphi=0.761$ (the 1989 observing season).

2. JD 2448810; $W=14.342^{m}$, $B=14.821^{m}$, $V=14.323^{m}$,
$R=14.068^{m}$ at $\varphi=0.861$ (the 1992 observing season).

 Here, the observer sees both components of the close binary in the
plane of the sky. Since the time scale is so short (exposure times
of 30--40~s in the $W$ and $B$~bands), we observe chaotic
variations with amplitudes up to $0.08^{m}{-}0.10^{m}$. It is
clear that such rapid variations testify to the presence of a
small radiating region, $(0.5{-}1.0)\times 10^{12}$~cm in size
(for any choice of model for the close binary).

 Our comparative analysis of the data in the $W$, $B$, $V$, and $R$
bands (Fig.~1) shows that the amplitude of the rapid variations is
very strongly dependent on wavelength, and increases with
decreasing wavelength, with the $W$ variations being largest and
sometimes reaching $0.05^{m}$ (Fig.~3). These rapid variations can
give rise to large scatter in the individual data points for the
dependences of $W{-}B$ on $B$ and of $B{-}V$ on $V$.

 It is natural to attribute the rapid variations of the binary to
considerable variability of the X-ray flux that heats the outer
regions of the accretion disk and a hot spot at the surface of the
optical component.

 The total amplitude of the system's irregular brightness
variations varies from $\Delta W \approx 1.15^m$ in the $W$~band
to $\Delta R\approx 0.85^m$ in the $R$~band. During the maximum
brightness, we observed chaotic variations with amplitudes from
0.085$^{m}$ to 0.190$^{m}$, with the rms uncertainty in the mean
brightness being ${\sim}0.005^{m}{-}0.007^{m}$ in the $W$ band. At
epochs of maximum brightness, the amplitude of rapid variations on
time scales of 60--90~s was ${\sim}0.03^{m}{-} 0.13^{m}$.

 The rms uncertainty in the mean in the $B$ and $V$ bands was
${\sim}0.005^{m}$ at maximum and ${\sim} 0.007^{m}$ at minimum
brightness (for observations with telescope apertures of at least
1000~mm). The strong scatter in the individual observational data
points is most likely due to the presence of rapid brightness
fluctuations during a single $WBVR$ exposure; the typical duration
of the $WBVR$ exposures was 8--10~minutes.

 Thus, the total amplitudes of the binary's rapid brightness
variations were approximately $\Delta W=1.35^m$, $\Delta
B=0.95^m$, $\Delta V=1.05^m$, and $\Delta R= 1.00^m$.

\section{COMPARISON OF THE OPTICAL OBSERVATIONS WITH THE X--RAY DATA FROM THE
EXOSAT AND Ginga OBSERVATORIES}

 X-ray data obtained for the low-mass binary Cyg~X-2 in 1987--1991
with the EXOSAT and Ginga observatories were analyzed in~[24]. The
three sectins, or branches, of the Z-shaped tracks were studied in
detail: the horizontal branch (HB), normal branch (NB), and
flaring branch (FB) (upper, middle, and lower parts of the
diagram, respectively).

 Notes that the system exhibited three states in the optical:
quiescence, the flaring state, and the active state, which is
intermediate between the first two states. The binary spends up to
several days in the intermediate state~[21, 22]. Optical
observations indicate that the system stays in the intermediate
state for two to days. The time interval for the binary's
intermediate state varies only slightly from one observing season
to another. A small amoung of rapid variability with an amplitude
of $\Delta = 0.3^m{-}0.4^m$ is also observed in quiescent state of
the system.

 The position along the Z-shaped track between the HB and FB is
related to variations in the rate at which matter is accreted onto
the neutron star~[4, 5], which are manifest in many photometric
features in the optical. For example, the observed oscillations in
the system can be explained by motions of gas flows at high
velocities in a strong gravitational field, which give rise to
oscillations reaching several percent of the total brightness from
the system~[34].

 It is noteworthy in studies of rapid variations in the system that
the quasi-periodic oscillations observed in the X-ray could also
make a considerable contribution to the optical light curves
(Figs.~1, 2).

 X-ray observations for time intervals when the close binary was in
its high or intermediate states (June 1987, June and October 1988,
November 1990, May--June 1991) are presented in~[24]. The
existence of two states of the system --- the high state and
intermediate state (for a short time) --- also follows from our
optical observations, in particular, those covering the time
intervals in question. A comparative analysis of the X-ray
observations for these time intervals was presented in~[7].

 There is a weak correlation between the optical and X-ray
observations, with the correlation coefficient being $10\%$. This
follows from an analysis of our optical observations and published
optical and X-ray data.

\section{BRIGHTNESS DEPENDENCE OF COLOR INDICES}

 Analyzing the binary's optical light curves for all the years of
our observations, we find firm evidence for a clear dependence of
the $W{-}B$, $B{-}V$, and $V{-}R$ color indices on the
brightnesses in the $B$, $V$, and $R$ bands (Fig.~5). All the
panels of Fig.~5 display rapid variations during this time, which
could lead to the observed strong scatter of the data points in
the light curves and color--magnitude diagrams (especially those
relating $B{-}V$ to $V$).

 The amplitude of the $W{-}B$ variations is about $0.42^{m}$. The
orbital variations of the maximum brightness, with amplitudes of
$\Delta W\approx 0.30^{m}$, $\Delta B\approx0.278^{m}$, $\Delta
V\approx0.265^{m}$, and $\Delta R\approx 0.26^{m}$ and clear
brightness minima at the two conjunctions, is clearly visble in
all the bands.

 The brightness minimum near phase $\varphi = 0.50$ is narrower: as
spectroscopic data show, the F~star passes through the periastron
of its orbit near the upper conjunction, and the time it spends at
these phases is small due to the eccentricity of the binary's
orbit.

 Another feature of the system is also noteworthy: we observe a
slight UV excess at both conjunctions of the close binary's
components, in both the quiescent and active phases. The $B{-}V$
color index demonstrates no functional dependence on the phase.

\section{EVOLUTIONARY STATUS AND THE POSITION OF
V1341~$\textrm{Cyg} = \textrm{Cyg}$~X-2 IN THE TWO-COLOR DIAGRAM}

Qualitative estimates of the position and evolutionary status of
the system using the $(W{-}B)$--$(B{-}V)$ diagram (Fig.~6) must
take into account the fact that most of the UV luminosity of
V1341~Cyg is probably due to re-radiated X-ray flux~[10, 17].
Thus, X-ray variations should give rise to optical variations on
the same time scales. This follows from a comparison of available
X-ray data and the optical data we have obtained over seven years.

In the re-radiation model, the soft component of the X-ray flux is
mainly re-radiated as optical photons and the X-ray radiation is
not isotropic. This also follows from several observational
results~[10] used for our analysis.

 Additional evidence supporting the re-radiation model is that the
radius of the re-radiating region adopted in our study, $R\sim
10^{12}$~cm (the accretion disk~$+$~the optical component) [35],
is close to the size of the binary itself (the component
separation is $d=1.7\times 10 ^{12}$~cm).

 The nightly-average color indices from our observations (corrected
for the adopted interstellar reddening, $E_{B{-}V}=0.22^{m}$ [36])
can be taken from Figs.~2, 6.

Note that, in its very lowest state, the optical component of
V1341~Cyg lies on the sequence of stars with normal metallicity,
and is located on the red part of the horizontal branch (Fig.~8).

The main features we would like to emphasize here (and which are
important for the entire study) are the following.

1. A very important conclusion drawn in~[5], which considerably
altered earlier distance estimates for Cyg~X-2 based on optical
observations (and used in~[12]) should be noted: the data in~[10]
differ by ${\sim} 30\%{-}60\%$ from those in~[5] (for example, the
distance given is $ d=8.7^{-1.8}_{+2.2} $).

2. The optical component of the close binary V1341~$\textrm{Cyg} =
\textrm{Cyg}$~X-2 is (for the adopted $E_{B{-}V}$) near the
main-sequence turn-off point for globular-cluster stars in the
evolutionary two-color diagram, and so is a dwarf or a star near
the turn-off to the red-giant branch (Fig.~8).

Taking into account these two features in the interpretation of
our seven-year optical observations and using the conclusions
of~[5] and~[12] when analyzing the color indices and comparing the
optical and X-ray data, we come into contradiction with the
suggestion that the primary component of the binary is a blue
straggler.

\section{CONTRIBUTION FROM THE ACCRETION DISK \protect\\ AND ACCRETION
STRUCTURES \protect\\ TO THE COMBINED LIGHT OF THE BINARY}

 To qualitatively estimate the contribution of the binary's
accretion disk to the total brightness, we used quantitative data
on the position of the object in the $(W{-}B){-}(B{-}V)$ two-color
diagram during different states of the system (Fig.~8a). Near the
primary minimum, Min~I, we observe rapid chaotic variations in the
$W$~band, with amplitudes between 0.085$^{m}$ and 0.190$^{m}$,
with uncertainties of ${\sim} 0.02^{m}$ (JD~2447414, JD~2447444).

 The primary minimum, Min~I, was observed (on average) near orbital
phase $\varphi=0.02$. During the 1986--1987 observations, Min~I
was observed near orbital phase $\varphi=0.01$. This ``floating''
of Min~I ($\varphi=0.01{-}0.02$) was detected in our observations
between 1986 and 1992.

 The binary's rapid, irregular variations are probably due to
variations in the X-ray flux heating the outer parts of the
accretion disk and the hot spot on the optical component, as well
as variations in the conditions for the matter flow from the
optical component through the inner Lagrange point, L$_{1}$.
Contrary to expectations, the earlier simultaneous observations
of~[23] were not able to detect correlations between variations of
the X-ray and optical fluxes~[37, 38].

 As was noted above, there is some flaring in the UV near orbital
phases $\varphi=0.3$ and $\varphi=0.8$ (Fig.~1), even when the
system is in quiescence. This high-temperature flaring (it is not
as prominent in the $V$ and $R$~bands!) can probably be
interpreted in the transition-layer model of~[5].

 This model considers the motion of a matter clump that is a source
of quasi-periodic oscillations (in this case, a ``blob''{} of
high-temperature plasma) on the accretion-disk surface, along a
Keplerian orbit around the neutron star. Let us assume that the
magnetosphere axis and the normal to the accretion-disk surface do
not coincide, forming a small angle $\delta$. Repeatedly passing
the slightly oblique magnetosphere, the blob is also influenced by
Coriolis forces that contribute to the disruptionn of the plasma
clump.

 The contribution of the accretion disk to the combined brightness
of the binary varies from close to $4\%{-}5 \%$ (quiescence) to
$50 \%$ (the active state).

\section{CONCLUSIONS}

 We have obtained a long-term, uniform set of observations in the
$W$, $B$, $V$, and $R$ bands, monitoring as long as possible on
each observing night. We have used this dataset to distinguish
fine photometric effects for the close binary V1341~$\textrm{Cyg}
= \textrm{Cyg}$~X\mbox{-}2.

 The optical light curves of V1341~$\textrm{Cyg} =
\textrm{Cyg}$~X-2 in quiescence during our 1986--1992 observations
are in qualitative agreement with the light curves of other
studies.

 Our observations indicate that the amplitude of the $B$-band
ellipsoidal brightness variations during the system's quiescence
is within $0.26^{m}{-} 0.28^{m}$~[21]. For comparison, the
amplitude of the ellipsoidal variations from the data of~[12] was
$\Delta B=0.27^{m}$.

 The depths of the minima in the $B$ and $V$~bands are the same
within the errors, ${\sim}$0.3$^{m}$. The $W$-band minima are
deeper by $0.015^{m}$, while those in the $R$~band are shallower
by $0.008^{m}$.

 The close binary V1341~$\textrm{Cyg} = \textrm{Cyg}$~X-2 exhibits
both chaotic variations and short-term optical flares, which are
probably due to activity of the X-ray source and gas flows in the
system.

 Simultaneous observations of V1341~$\textrm{Cyg} =
\textrm{Cyg}$~X-2 in the X-ray and optical prior to 1987 did not
indicate any correlations between brightness variations in the
optical and X-ray~[39], while such correlations episodically
appeared between 1987 and 1991~[7].

 Instabilities in the disk accretion or the outer structures in the
accretion disk can give rise to changes in the flux gradients in
the X-ray and optical, and so an absence of correlations between
the optical and X-ray brightness variations. Like the observations
of other studies, our observations for the 1986 and 1992 observing
seasons partially support this possibility.

 Observations of the variable with integration times of 60--90~s
(in each band) indicate rapid, irregular brightness variations
with amplitudes ${\sim}0.04^{m}{-}0.05^{m}$ in the UV and somewhat
lower in the other bands.

 The system demonstrates the largest-amplitude variations
(${\sim}1.25^{m}$) in the UV in its active stage (the 1987 and
1989 observing seasons); at that time, the brightness varied
between $14.15^{m}$ and $15.40^{m}$.

 The light curves for the 1986--1992 observing seasons (at certain
epochs) reveal a narrow minimum of the binary light curve near the
phase of the lower conjunction ($\varphi = 0.50 $), in both
quiescence and the active state. This somewhat exceeds the primary
minimum in its amplitude (Fig.~1).

 We confirm the presence of rapid, irregular variations on
timescales of about 60--90~s, with the $W$-band amplitude from
$0.085^{m}$ to $0.190^{m}$ and uncertainties
${\sim}0.005^{m}{-}0.007^{m}$. The variations in the other bands
are between $0.065^{m}$ and $0.140^{m}$, with uncertainties of
${\sim} 0.005^{m}$ (for telescope apertures of 1000~mm or more).

 Our comparative analysis of our $WBVR$ data demonstrates that the
amplitudes of the rapid variations strongly depend on the
wavelength: they grow with decreasing wavelength and are largest
in the $W$~band. These rapid, irregular variations are probably
due to changes in the X-ray flux heating the outer layers of the
accretion disk and the hot spot on the optical component, as well
as to conditions for the matter flow from the optical component
through the inner Lagrange point L$_{1}$.

 A considerable contribution to the total UV light leaving the
system also comes from plasma clumps (``blobs'') whose presence is
manifest at optical phases $\varphi=0.3$ and 0.8. The brightnesses
of such luminous plasma clumps are from ${\sim}0.50^{m}$ to
${\sim}0.90^{m}$, with uncertainties of ${\sim}0.01
^{m}{-}0.02^{m}$ in the $W$ and $B$ bands, respectively.

\begin{figure*}[t!]
\includegraphics{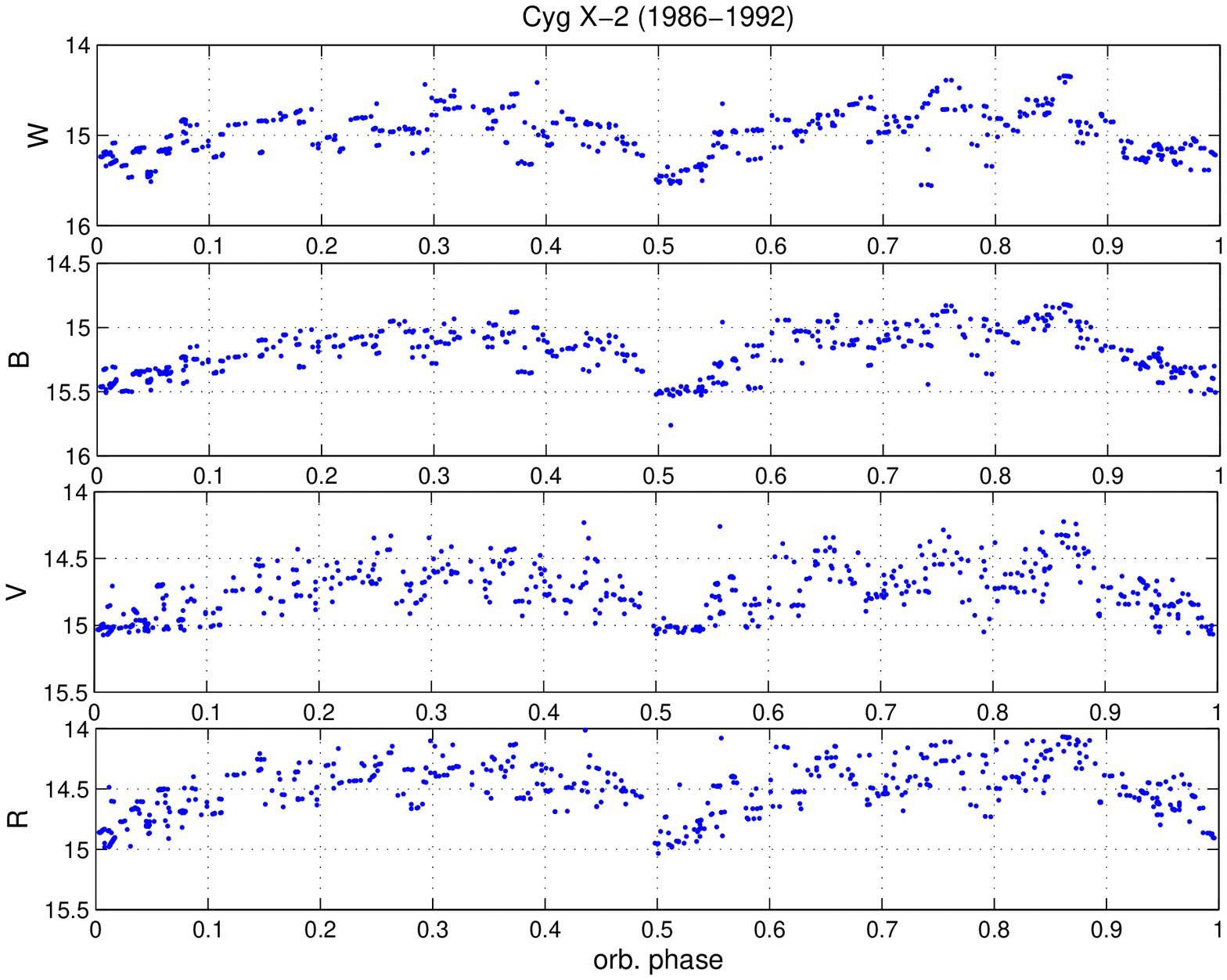}
\centerline{\epsfig{file=Sazonov1.eps,width=100mm}}
\caption{$WBVR$ light curves of V1341~Cyg~$=$~Cyg~X-2 folded with
the orbital period $P=9.8431^{\textrm{d}}$. ($\textrm{Min JD} =
2443161.7 + 9.8431 E$.) The object was observed in 1986--1992.
\hfill}
\end{figure*}

\begin{figure*}[t!]
\includegraphics{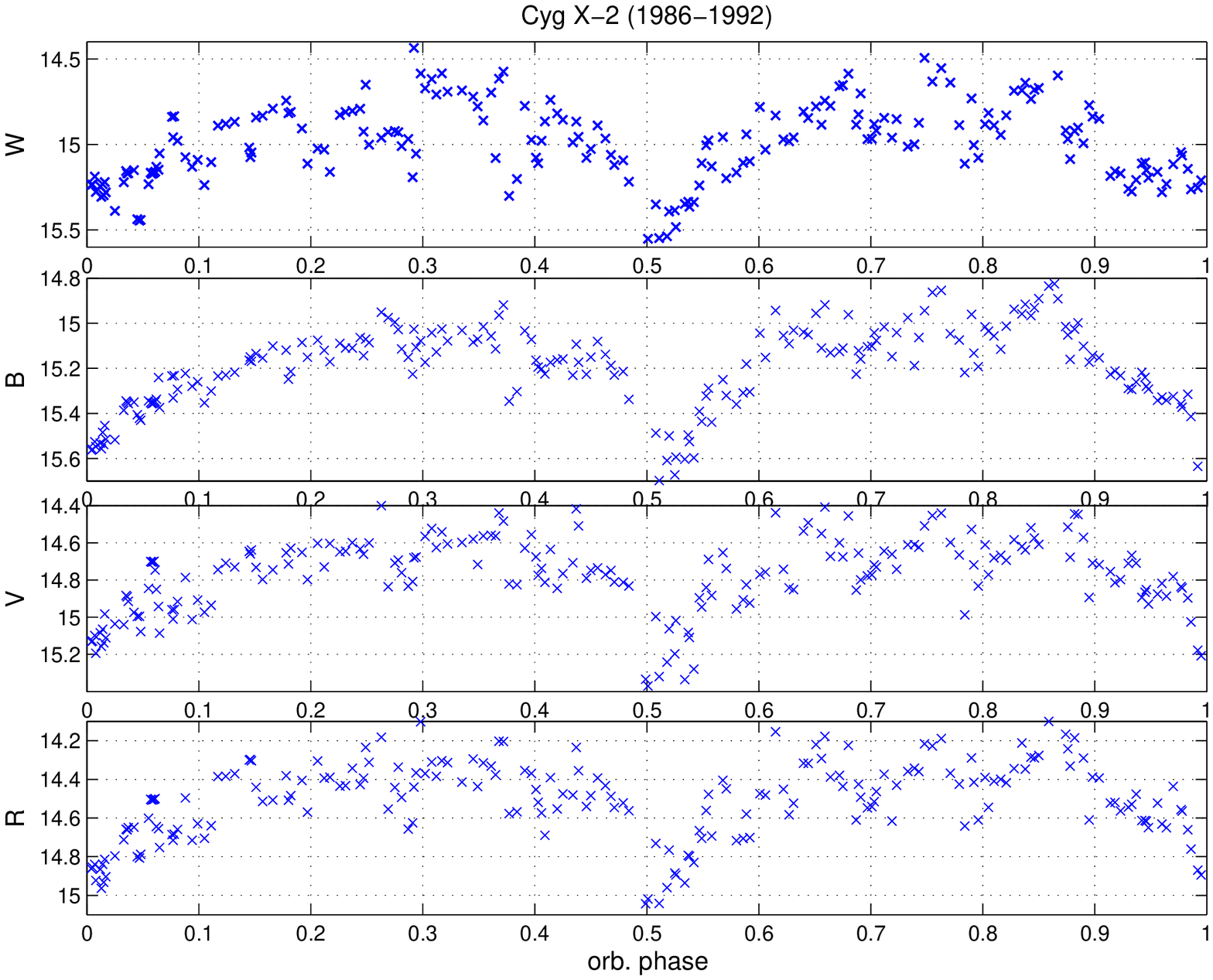}
\centerline{\epsfig{file=Sazonov2.eps,width=100mm}}
\caption{Mean light curves of
V1341~$\textrm{Cyg}=\textrm{Cyg}$~X-2 in the $WBVR$ bands, plotted
for the system in quiescence and averaged within
orbital-phase bins. The observations for 1986--1992 are shown. 
\hfill}
\end{figure*}

\begin{figure*}[t!]
\includegraphics{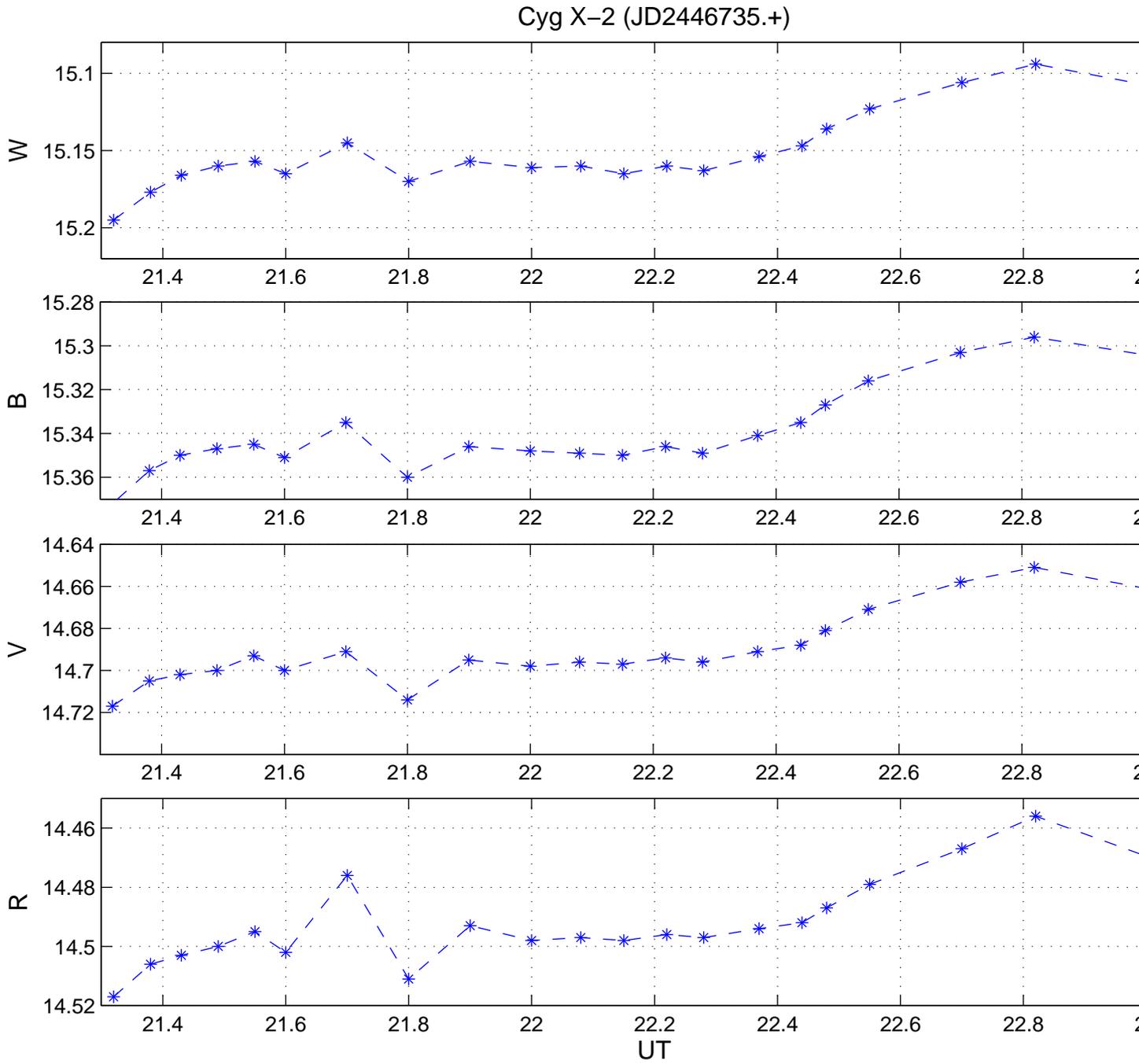}
\caption{$WBVR$ light curves of V1341~$\textrm{Cyg} =
\textrm{Cyg}$~X-2
during a rapid flare of the second type on JD~2446735. 
\hfill}
\end{figure*}

\begin{figure*}[t!]
\includegraphics{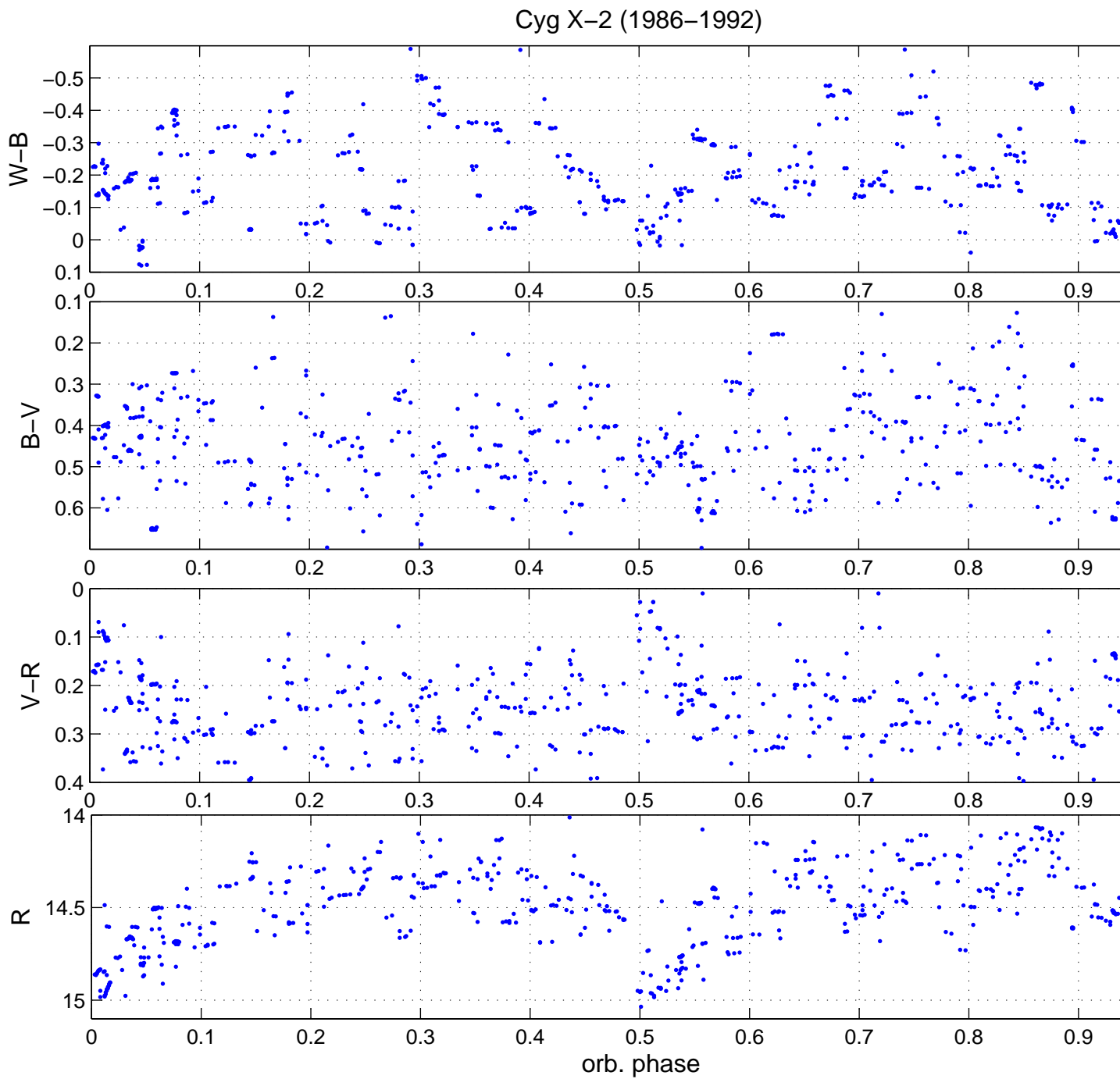}
\caption{The $W{-}B$, $B{-}V$, $V{-}R$ color indices averaged in
phase bins and the $R$-band light curve of V1341~$\textrm{Cyg} =
\textrm{Cyg}$~X-2 from
observations in 1986--1992. 
\hfill}
\end{figure*}

\begin{figure*}[t!]
\includegraphics{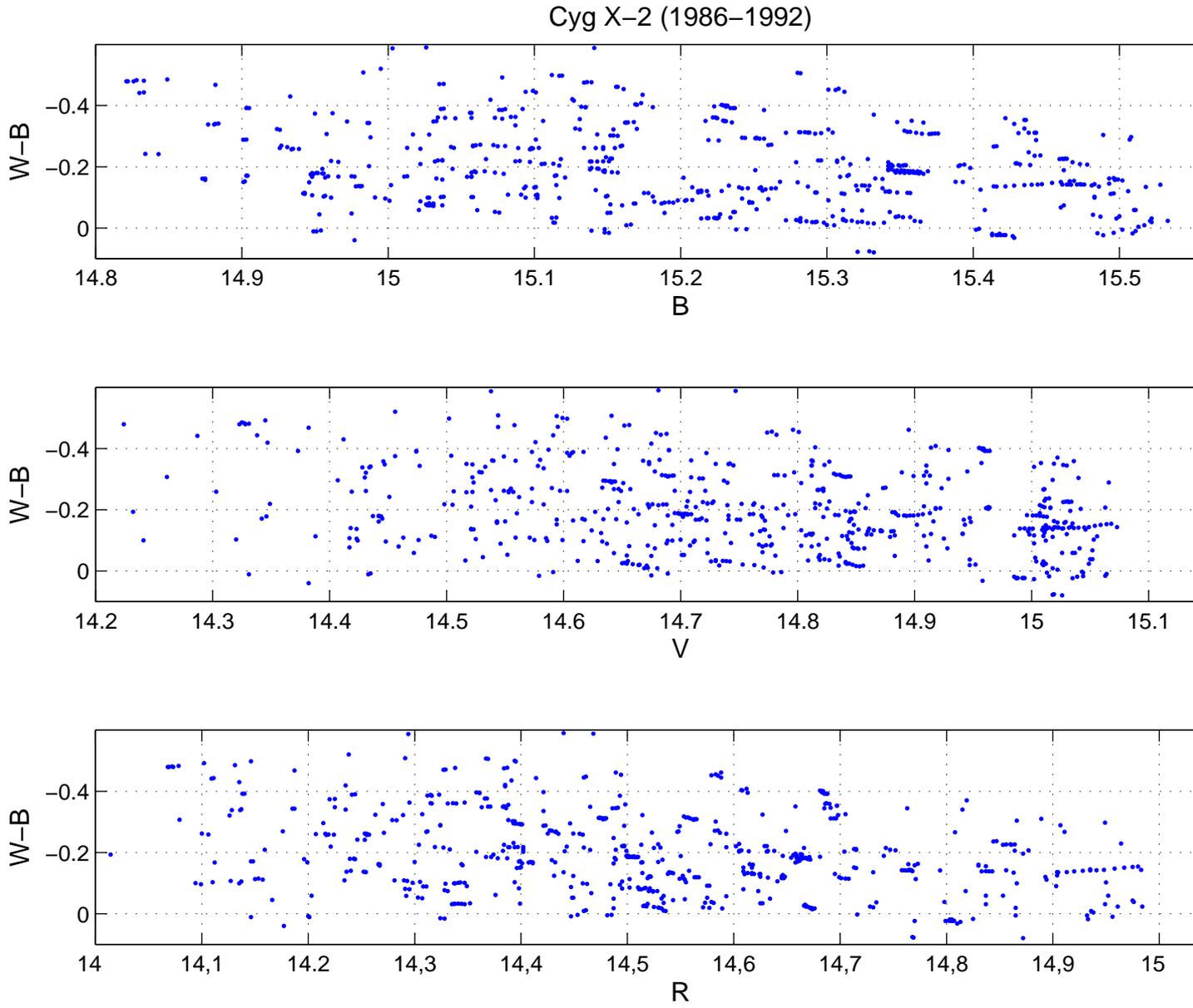}
\vspace*{10pt}
\caption{The (a) $W{-}B$, (b) $B{-}V$, and (c)
$V{-}R$ color indices versus the $B$, $V$, and $R$~magnitudes
[only $V$ and $R$ for panel (c)], from observations in
1986--1992.
\hfill}
\end{figure*}

\begin{figure*}[t!]
\includegraphics{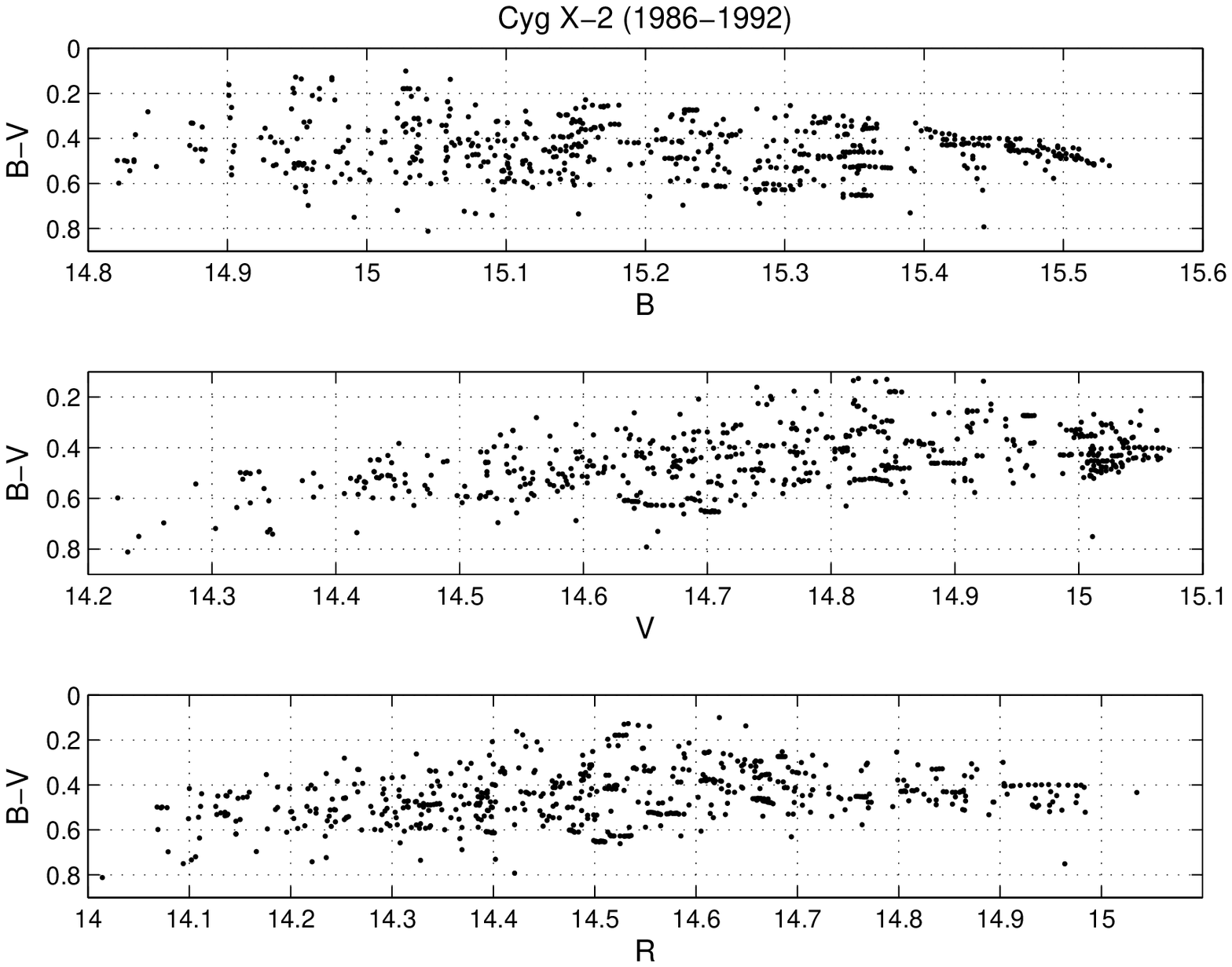}
\addtocounter{figure}{-1}
\vspace*{10pt}
\caption{\textbf{~~~~~~~~~~~~~~~~~~~~~~~~~~~~~~~~~~~~~~~~~~~~~~~~~~Fig.5.}
(Continued.)
\hfill}
\end{figure*}

\begin{figure*}[t!]
\includegraphics{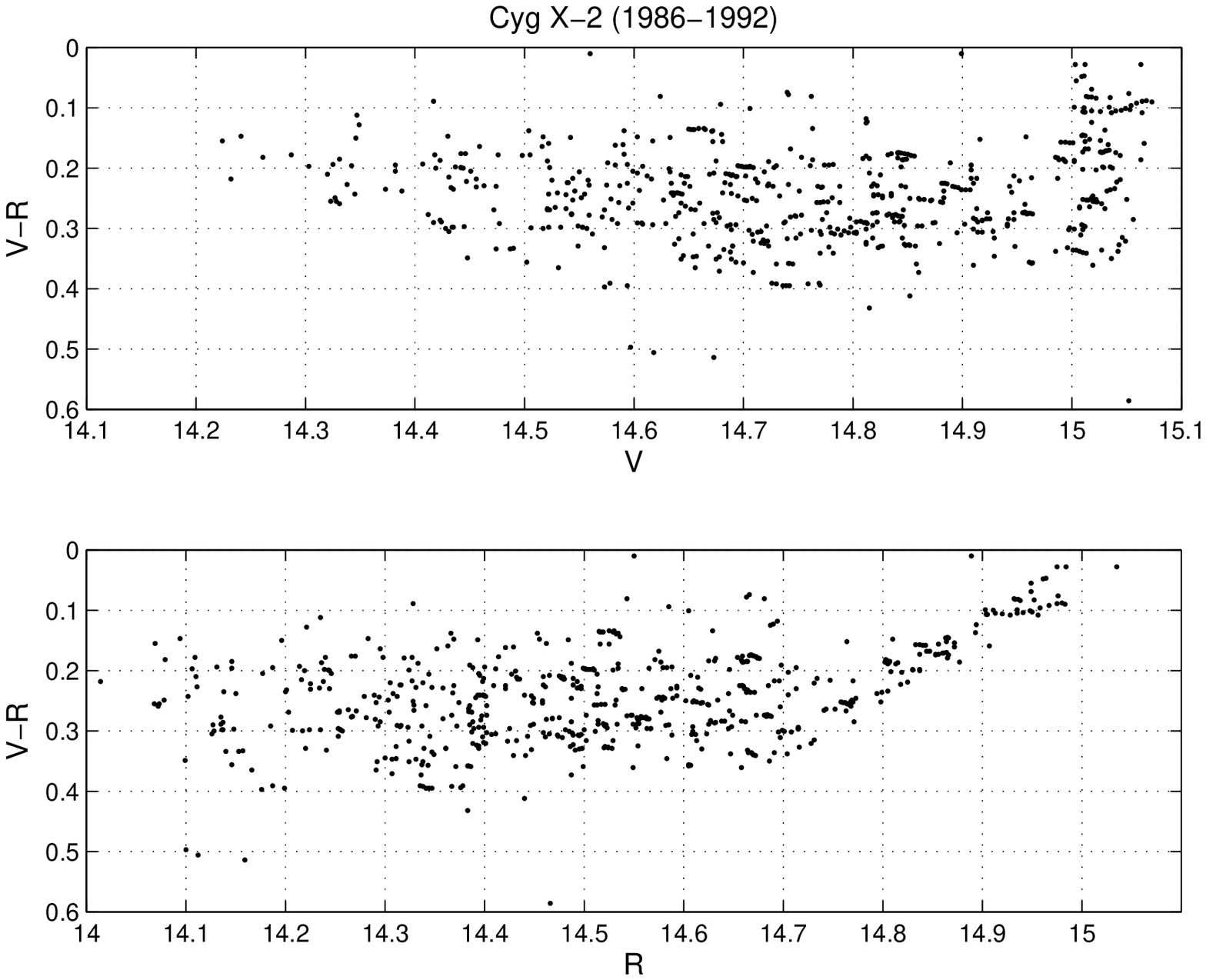}
\addtocounter{figure}{-1}
\vspace*{10pt}
\caption{\textbf{~~~~~~~~~~~~~~~~~~~~~~~~~~~~~~~~~~~~~~~~~~~~~~~~~~Fig.
5.} (Continued.)
\hfill}
\end{figure*}

\begin{figure*}[t!]
\includegraphics{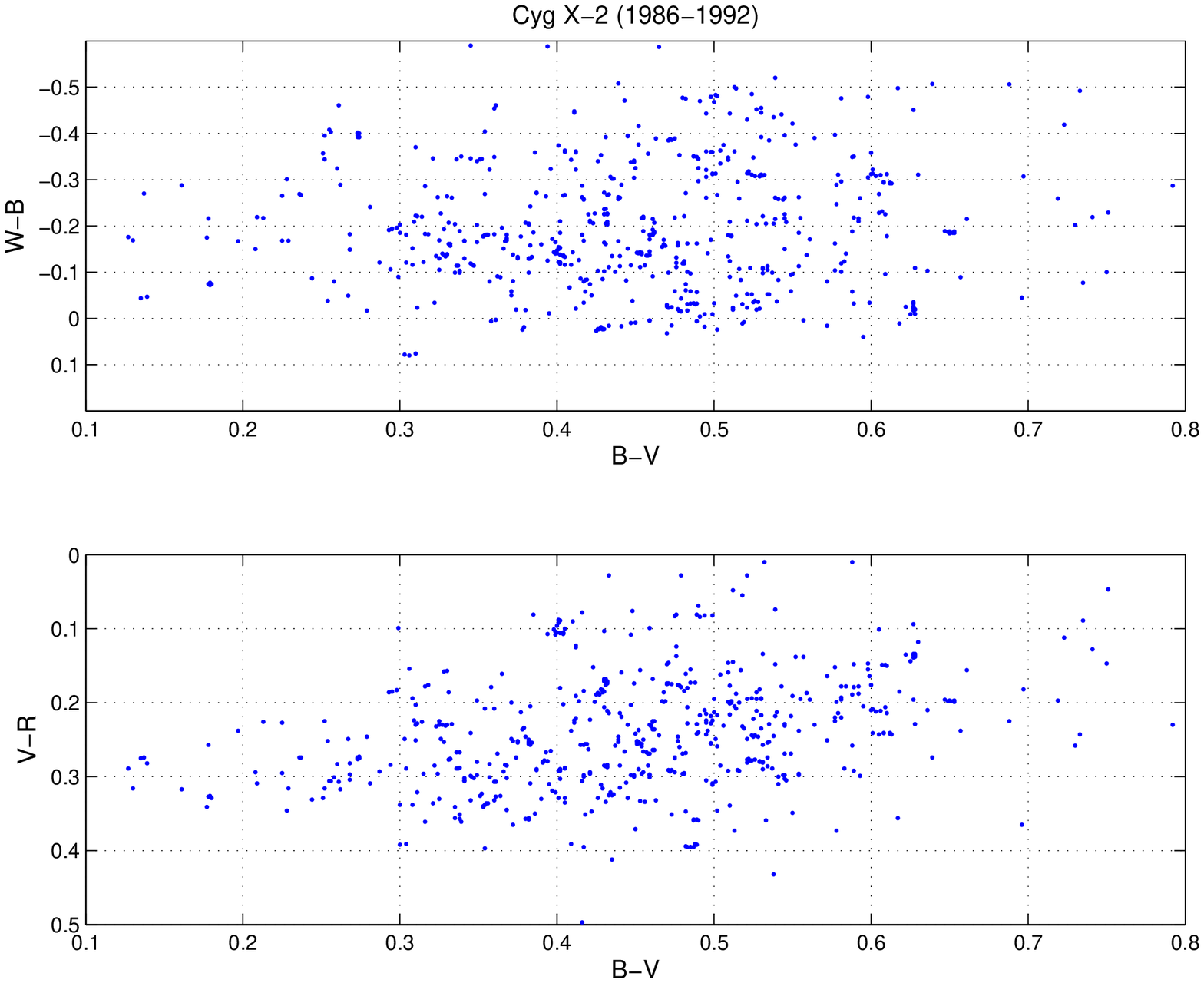}
\vspace*{10pt}
\caption{The ($W{-}B$)--($B{-}V$)~(top) and
($V{-}R$)--($B{-}V$)~(bottom) two-color diagrams for
V1341~Cyg~$=$~Cyg~X-2, from observations in 1986--1992.
\hfill}
\end{figure*}

\begin{figure*}[p!]
\includegraphics{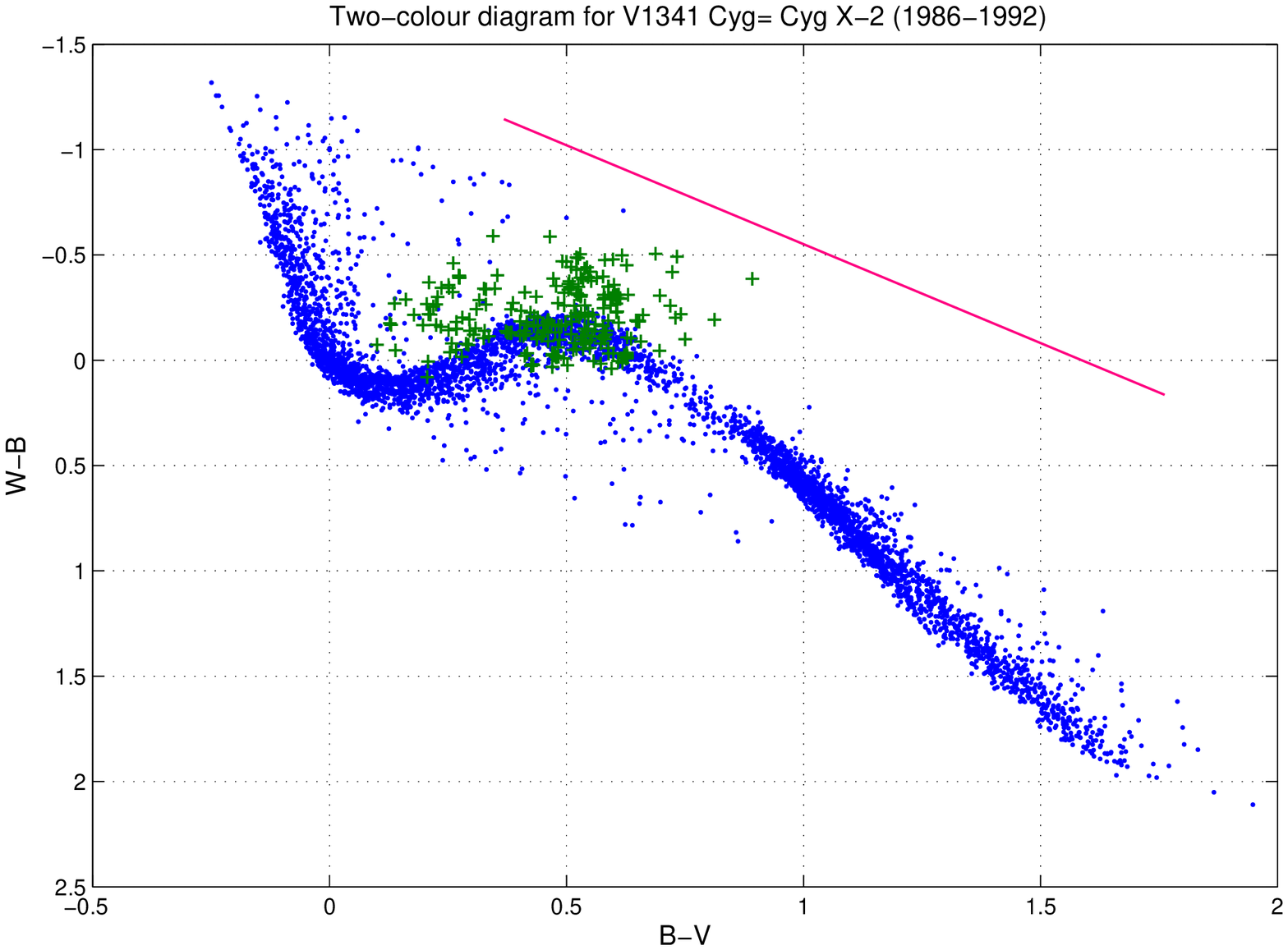}
\vspace*{10pt}
\caption{Positions of V1341~$\textrm{Cyg}
=\textrm{Cyg}$~X-2 in the (a) ($W{-}B$)--($B{-}V$) and (b)
($U{-}B$)--($B{-}V$) two-color diagrams for various activity
levels (crosses), compared to those for globular-cluster stars and
cataclysmic variables (dots). The interstellar reddening line is
plotted as a solid line. The dashed curve in panel~(a) corresponds
to the positions of blackbodies of different temperatures. The
circles are the mean positions of V1341~$\textrm{Cyg} =
\textrm{Cyg}$~X-2 in its high (\emph{1}), intermediate (\emph{2}),
and low (\emph{3}) states. The Alma-Ata $WBVR$ photometric system
is used. The $U{-}B$ values in panel~(b) are from the Alma-Ata
catalog [31, 32]. \hfill}
\end{figure*}

\begin{figure*}[p!]
\includegraphics{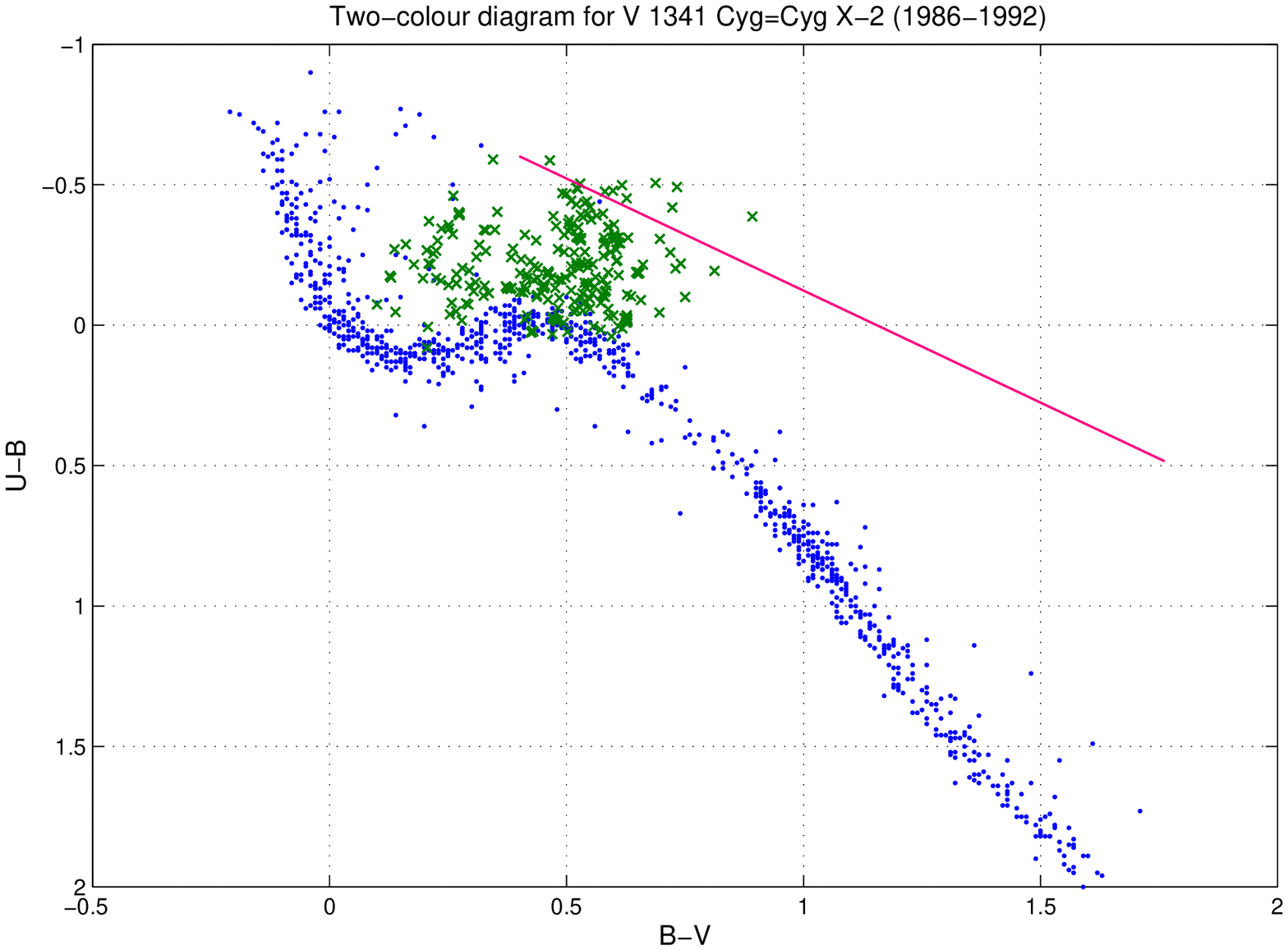}
\vspace*{10pt}
\caption{Positions of V1341~$\textrm{Cyg}
=\textrm{Cyg}$~X-2 in the (a) ($W{-}B$)--($B{-}V$) and (b)
($U{-}B$)--($B{-}V$) two-color diagrams for various activity
levels (crosses), compared to those for globular-cluster stars and
cataclysmic variables (dots). The interstellar reddening line is
plotted as a solid line. The dashed curve in panel~(a) corresponds
to the positions of blackbodies of different temperatures. The
circles are the mean positions of V1341~$\textrm{Cyg} =
\textrm{Cyg}$~X-2 in its high (\emph{1}), intermediate (\emph{2}),
and low (\emph{3}) states. The Alma-Ata $WBVR$ photometric system
is used. The $U{-}B$ values in panel~(b) are from the Alma-Ata
catalog [31, 32].
\hfill}
\end{figure*}


\newpage

\begin{thebibliography}{99}

\addtolength{\itemsep}{-1.5pt}
\bibitem{Giacconi1967:Sazonov_n} R.~Giacconi, P.~Gorenstein, H.~Gursky, et al., Astrophys. J. \textbf{148}, L129 (1967).
\bibitem{Hasinger1989:Sazonov_n} G.~Hasinger and M.~Van~der~Klis, Astron. Astrophys. \textbf{225}, 79 (1989).
\bibitem{Lamb1991:Sazonov_n} F.~K.~Lamb, in \emph{Neutron Stars: Theory and Observations}, Ed. by J.~Ventura and D.~Pines, NATO ASI Ser. C, v.~344 (Dordrecht: Kluwer, 1991), p.~445.
\bibitem{Kuznezov2001:Sazonov_n} S.~I.~Kuznetsov, Pis'ma Astron. Zh. \textbf{27}, 919 (2001) Astron. Lett. \textbf{27}, 790 (2001).
\bibitem{Kuznezov2002:Sazonov_n} S.~I.~Kuznetsov, Pis'ma Astron. Zh. \textbf{28}, 88 (2002) Astron. Lett. \textbf{28}, 73 (2002).
\bibitem{Kuulkers1996:Sazonov_n} E.~Kuulkers and M.~van~der~Klis, Astron. Astrophys. \textbf{314}, 567 (1996).
\bibitem{Kuulkers1996a:Sazonov_n} E.~Kuulkers, M.~van~der~Klis, and B.~A.~Vaughan, Astron. Astrophys. \textbf{311}, 197 (1996).
\bibitem{Kuulkers1994a:Sazonov_n} E.~Kuulkers, M.~van~der~Klis, T.~Oosterbroek, et al., Astron. Astrophys. \textbf{289}, 795 (1994).
\bibitem{Kuulkers1996b:Sazonov_n} E.~Kuulkers, M.~van~der~Klis, T.~Oosterbroek, et al., Mon. Not. R.~Astron. Soc. \textbf{287}, 495 (1997).
\bibitem{Cowley1979:Sazonov_n} A.~P.~Cowley, D.~Crampton, and J.~B.~Hutchings, Astrophys. J. \textbf{231}, 539 (1979).
\bibitem{Crampton1980:Sazonov_n} D.~Crampton and A.~P.~Cowley, Publ. Astron. Soc. Pacif. \textbf{92}, 147 (1980).
\bibitem{Goranskii1988:Sazonov_n} V.~P.~Goransky and V.~M.~Lyuty, Astron. Zh. \textbf{65}, 385 (1988) [Sov. Astron. \textbf{32}, 193 (1988)].
\bibitem{Vritlek1986:Sazonov_n} S.~D.~Vrtilek, S.~M.~Kahn, L.~E.~Grindlay, et al., Astrophys. J.~\textbf{307}, 698 (1986).
\bibitem{Vritlek1988:Sazonov_n} S.~D.~Vrtilek, J.~H.~Swank, R.~L.~Kelly, and S.~M.~Kahn, Astrophys. J.~\textbf{329}, 276 (1988).
\bibitem{Hasinger1988:Sazonov_n} G.~Hasinger, in \emph{Physics of Neutron Stars and Black Holes}, Ed. by Y.~Tanaka (Universal Acad., Tokyo, 1988), p.~97.
\bibitem{Hasinger1990:Sazonov_n} G.~Hasinger, M.~van~der~Klis, K.~Ebisawa, et al., Astron. Astrophys. \textbf{235}, 131 (1990)
\bibitem{Luyty1976:Sazonov_n} V.~M.~Lyuty and R.~A.~Syunyaev, Astron. Zh. \textbf{53}, 511 (1976) [Sov. Astron. \textbf{20}, 290 (1976)].
\bibitem{Smale1998:Sazonov_n} A.~Smale, Astrophys. J. \textbf{498}, L141 (1998).
\bibitem{Miller1998:Sazonov_n} M.~C.~Miller, F.~K.~Lamb, and D.~Psaltis, Astrophys. J. \textbf{508}, 791 (1998).
\bibitem{Strayzhis1977:Sazonov_n} V.~L.~Straizys, \emph{Multicolor Stellar Photometry} (Mosklas, Vil'nyus, 1977; Pachart Publ., Tucson, 1992).

\bibitem{Sazonov1988:Sazonov_n} A.~N.~Sazonov, Astron. tsirk. \textbf{1531}, 15 (1988).

\bibitem{Sazonov2006:Sazonov_n} A.~N.~Sazonov, in \emph{Proc. of the All-Russ. Astron. Conf. on Close Binary Stars in Modern Astrophysics, Moscow, Sternberg Astron. Inst., 22--24 May, 2006} (GAISh, Moscow, 2006), p.~39.
\bibitem{Ilovaisky1978:Sazonov_n} S.~A.~Ilovaisky, C.~Chevalier, M.~Chevreton, and S.~Bonazzola, Astron. Astrophys. \textbf{67}, 287 (1978).
\bibitem{Wijnands1997:Sazonov_n} R.~A.~D.~Wijnands, M.~van~der~Klis, E.~Kuulkers, et al., Astron. Astrophys. \textbf{323}, 399 (1997).
\bibitem{Berdnikov1986:Sazonov_n} L.~N.~Berdnikov, Perem. zvezdy \textbf{22}, 369 (1986).
\bibitem{Moshkalev1985:Sazonov_n} V.~G.~Moshkalev and Kh.~F.~Khaliullin, Astron. Zh. \textbf{62}, 393 (1985) [Sov. Astron. \textbf{29}, 227 (1985)].
\bibitem{Moffett1979:Sazonov_n} T.~J.~Moffett and G.~T.~Barnes, Astron. J.~\textbf{84}, 627 (1979).
\bibitem{Zakharov2006:Sazonov_n} A.~I.~Zakharov and A.~V.~Mironov, private commun. (2006).
\bibitem{Kornilov1991:Sazonov_n} V.~G.~Kornilov, I.~M.~Volkov, A.~I.~Zakharov et al., Tr. Gos. Astron. In-ta \textbf{63}, 1 (1991).
\bibitem{Harmanec1998:Sazonov_n} P.~Harmanec, Astron. Astrophys. \textbf{335}, 173 (1998).
\bibitem{Mironov2002:Sazonov_n} A.~V.~Mironov and A.~I.~Zakharov, Astrophys. Spase Sci. \textbf{280}, 71 (2002).
\bibitem{Zakharov2004:Sazonov_n} A.~V.~Zakharov, A.~V.~Mironov, and A.~N.~Krutyakov, Tr. Gos. Astron. Inst. \textbf{70}, 289 (2004).
\bibitem{Crampton1978:Sazonov_n} D.~Crampton and A.~P.~Cowley, IAU Circ. No.~3292 (1978).
\bibitem{Hasinger1987:Sazonov_n} G.~Hasinger, in \emph{The Origin and Evolution of Neutron stars}, Ed. by D.~J.~Helfand and Z.~Huang, IAU Symp. \textbf{125}, 333 (1987).
\bibitem{Beskin1979:Sazonov_n} G.~M.~Beskin, S.~I.~Neizvestnyi, A.~A.~Pimonov, et al., Pis'ma Astron. Zh. \textbf{5}, 271 (1979) [Sov. Astron. Lett. \textbf{5}, 144 (1979)].
\bibitem{Cathey1976:Sazonov_n} L.~R.~Cathey and J.~E.~Hayes, Astrophys. J. \textbf{151}, L89 (1976).
\bibitem{Branduardi-Raymont1984:Sazonov_n} G.~Branduardi-Raymont, L.~Chiapetti, E.~N.~Ercan, Astron. Astrophys. \textbf{130}, 175 (1984).
\bibitem{Hasinger1985:Sazonov_n} G.~Hasinger, A.~Langmeier, M.~Sztajno, and N.~White, IAU Circ. No. 4070, (1985).
\bibitem{Ilovaisky1979:Sazonov_n} S.~A.~Ilovaisky, C.~Chevalier, C.~Motch, and E.~Janot-Pachrco, IAU Circ. No.~3325 (1979).





\newpage
\end{thebibliography}
\end{document}